\documentclass[12pt]{article}
\usepackage{graphicx}
\usepackage[english]{babel}
\usepackage{amsmath,amsthm}
\usepackage{amsfonts}
\usepackage{lineno}
\usepackage{mathrsfs}
\usepackage{geometry}
\usepackage{amssymb}
\usepackage{amsmath}
\usepackage{booktabs}
\usepackage{enumerate}
\usepackage{verbatim}
\usepackage{graphicx}
\usepackage{epsfig}
\usepackage{epsf}
\usepackage{epstopdf}
\usepackage{amstext}
\usepackage{latexsym}
\usepackage{subfigure}
\usepackage[matrix,frame,arrow]{xypic}
\usepackage{makeidx}
\usepackage{CJK}
\usepackage{cite}

\newcommand\pubnumber{DPF2013-93}
\newcommand\pubdate{\today}

\def\napoli{Department of Physics and Astronomy\\
University of New Mexico, Albuquerque, NM, USA, 87131}
\def\support{\footnote{This work was supported by the US Department of Energy.}}

\def\Title#1{\begin{center} {\Large #1 } \end{center}}
\def\Author#1{\begin{center}{ \sc #1} \end{center}}
\def\Address#1{\begin{center}{ \it #1} \end{center}}

\newcommand\pubblock{\rightline{\begin{tabular}{l} \pubnumber\\
         \pubdate  \end{tabular}}}
\newenvironment{Abstract}{\begin{quotation}  }{\end{quotation}}
\newenvironment{Presented}{\begin{quotation} \begin{center}
             PRESENTED AT\end{center}\bigskip
      \begin{center}\begin{large}}{\end{large}\end{center} \end{quotation}}

\def\beq{\begin{equation}}
\def\eeq#1{\label{#1}\end{equation}}
\def\eeqn{\end{equation}}

\def\beqa{\begin{eqnarray}}
\def\eeqa#1{\label{#1}\end{eqnarray}}
\def\eeqan{\end{eqnarray}}

\let\bar=\overbar

\def\eg{{\it e.g.}}

\def\L{{\cal L}}

\def\Dslash{\not{\hbox{\kern-4pt $D$}}}
\def\dslash{\not{\hbox{\kern-2pt $\del$}}}

\def\msb{{\bar{\ssstyle M \kern -1pt S}}}

\begin{document}
\begin{titlepage}
\pubblock

\vfill
\Title{Production and spectroscopy of hadrons containing a b quark at ATLAS}
\vfill
\Author{ Rui Wang\support}
\Address{\napoli}
\Author{ On behalf of the ATLAS Collaboration.}
\vfill
\begin{Abstract}
We present studies of the production and spectroscopy of some members of the B-hadron family. We
reconstruct B ground states in their hadronic decay modes with a $J/\psi$ or $\Upsilon$ in the final
state. These studies are based on the 2011 7 TeV dataset collected by the ATLAS detector.
\end{Abstract}
\vfill
\begin{Presented}
DPF 2013\\
The Meeting of the American Physical Society\\
Division of Particles and Fields\\
Santa Cruz, California, August 13--17, 2013\\
\end{Presented}
\vfill
\end{titlepage}
\def\thefootnote{\fnsymbol{footnote}}
\setcounter{footnote}{0}
%
\section{Introduction}
The production of B-hadrons at the Large Hadron Collider (LHC) provides particular challenges and opportunities for insight into Quantum Chromodynamics (QCD), especially as a probe of the properties of the fundamental constituents of matter and their interactions in a new energy regime. Mesons and baryons containing a b quark can be seen as the hydrogen and helium atoms of QCD, therefore the measurement of their spectra plays an especially important role in understanding the strong interactions. The characteristices of b-hadrons at the LHC can also give input for tuning models and refining event generators in a new energy regime.\\

The most important elements of the ATLAS detector for B physics measurements are the Inner Detector (ID) tracker, the Electromagnetic Calorimeter, and the Muon Spectrometer (MS); details can be found in \cite{detector}. Dedicated B physics triggers are based on both single muons and di-muons with different thresholds and mass ranges.

\section{Measurement of the $B^+$ production cross-section}
\begin{figure}[!htb]
  \centering
  \includegraphics[height=2in]{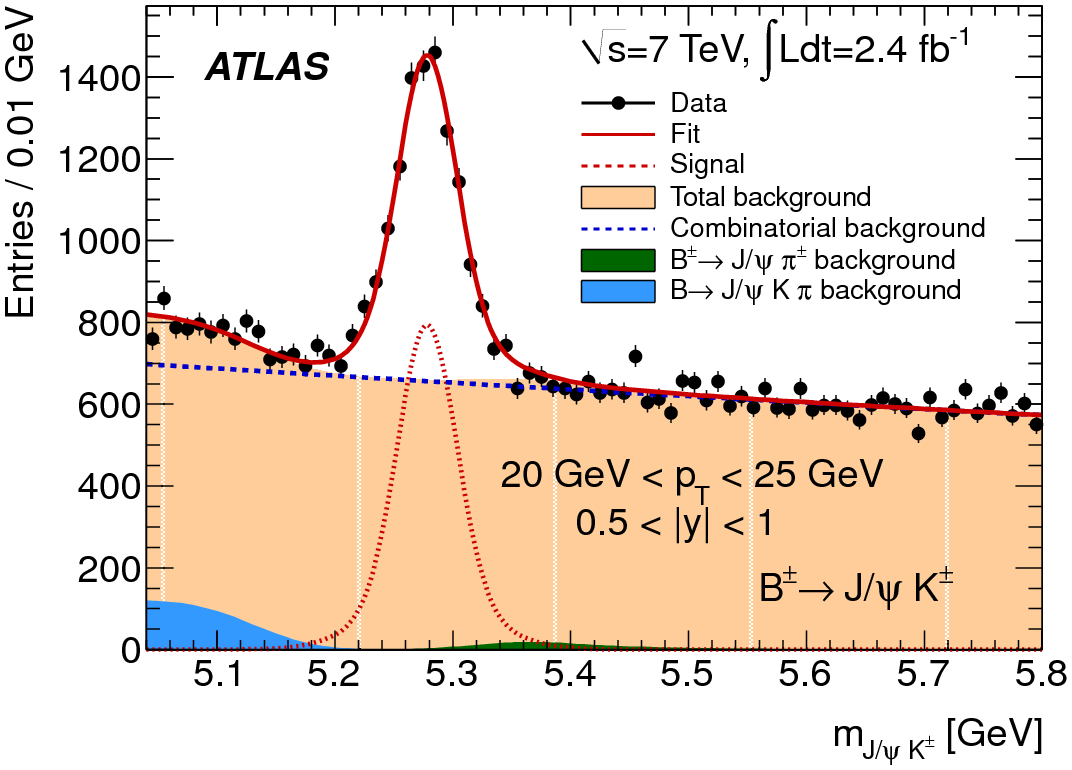}
  \caption{ The observed invariant mass distribution of $B^\pm$ candidates, $m_{J/\psi K^\pm}$, with transverse momentum and rapidity in the range 20 GeV $ <p_T < $ 25 GeV, 0.5 $ < |y| < $ 1 (dots), compared to the binned maximum likelihood fit (solid line). The error bars represent the statistical uncertainty. Also shown are the components of the fit as described in the legend.}
  \label{fig:bplusfit}
\end{figure}
Several B-hadron production cross sections have been measured \cite{bplus1,bplus2,bplus3,bplus4,bplus5,bplus6,bplus7,bplus8,bplus9,bplus10,bplus11,bplus12}, but the theoretical uncertainty remains up to 40\% due to uncertainty in the factorization and renormalization scales of the b-quark production, and the b-quark mass $m_b$ \cite{mb}. The LHC provides an opportunity to perform more precise measurements of B-hadron production cross-sections. Also the B-hadron production cross-section measurements in pp collisions at the LHC can provide tests of QCD calculations for heavy-quark production at high center-of-mass energies and in wide transverse momentum ($p_T$) and rapidity ($y$) ranges.\\
\begin{figure}[!htb]
  \centering
        \subfigure[]{\label{a}\includegraphics[height=2in]{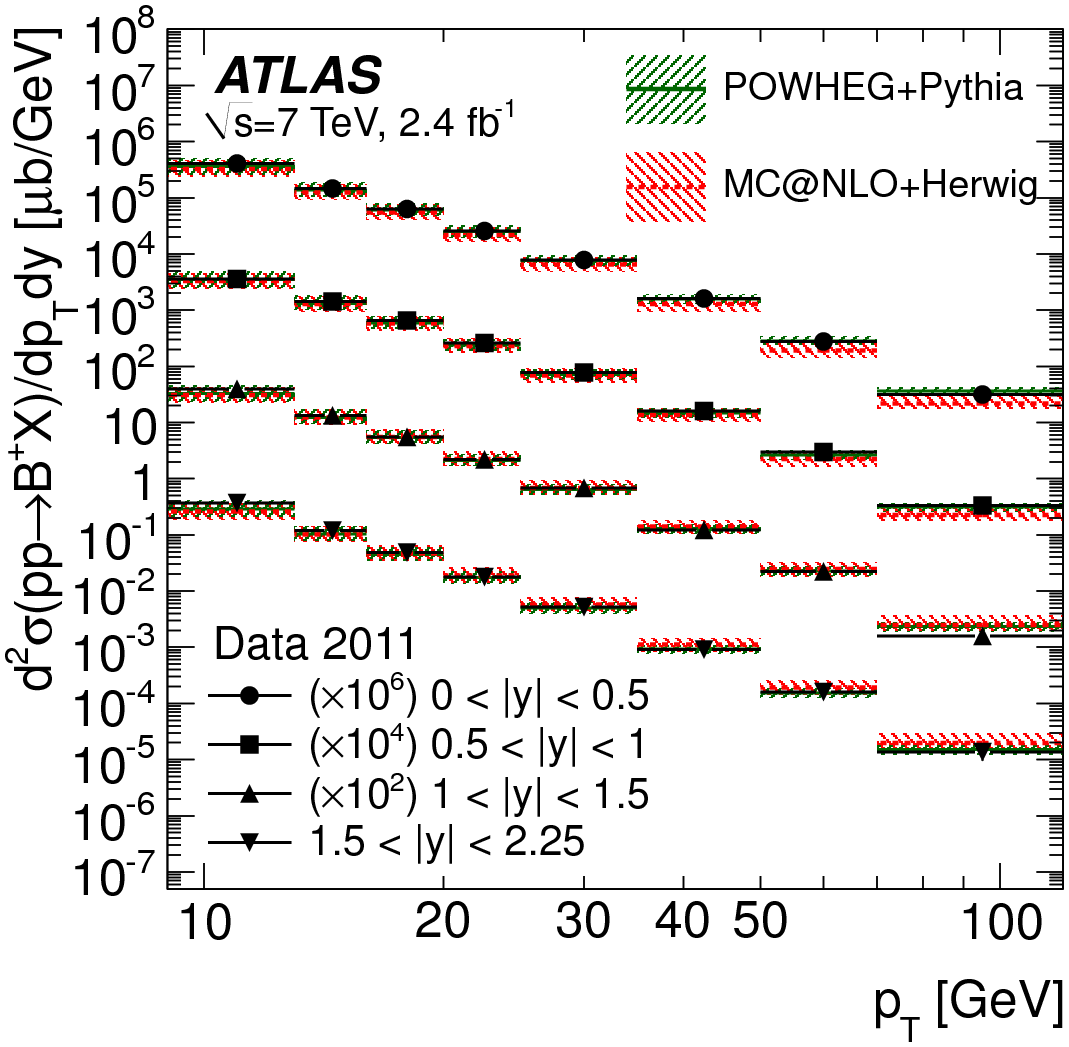}}
    	\subfigure[]{\label{b}\includegraphics[height=2in]{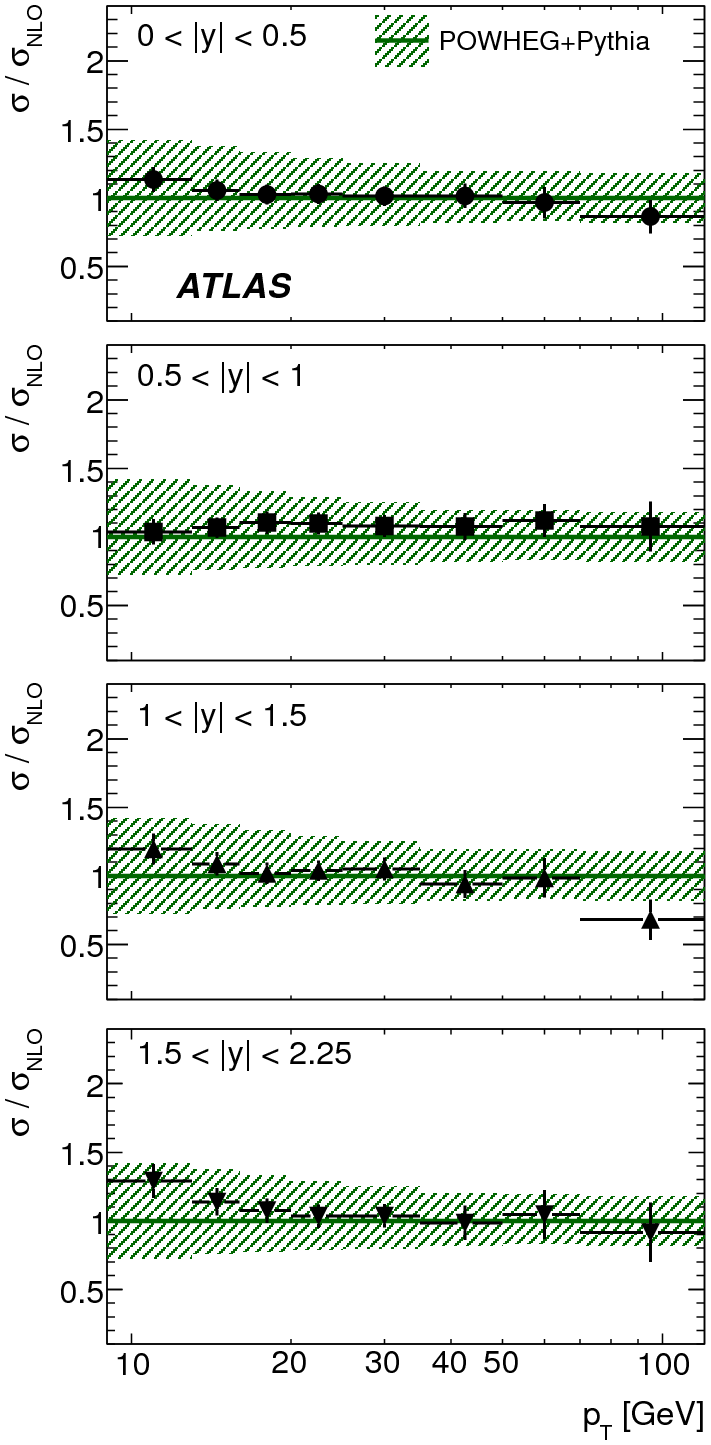}}
    	\subfigure[]{\label{c}\includegraphics[height=2in]{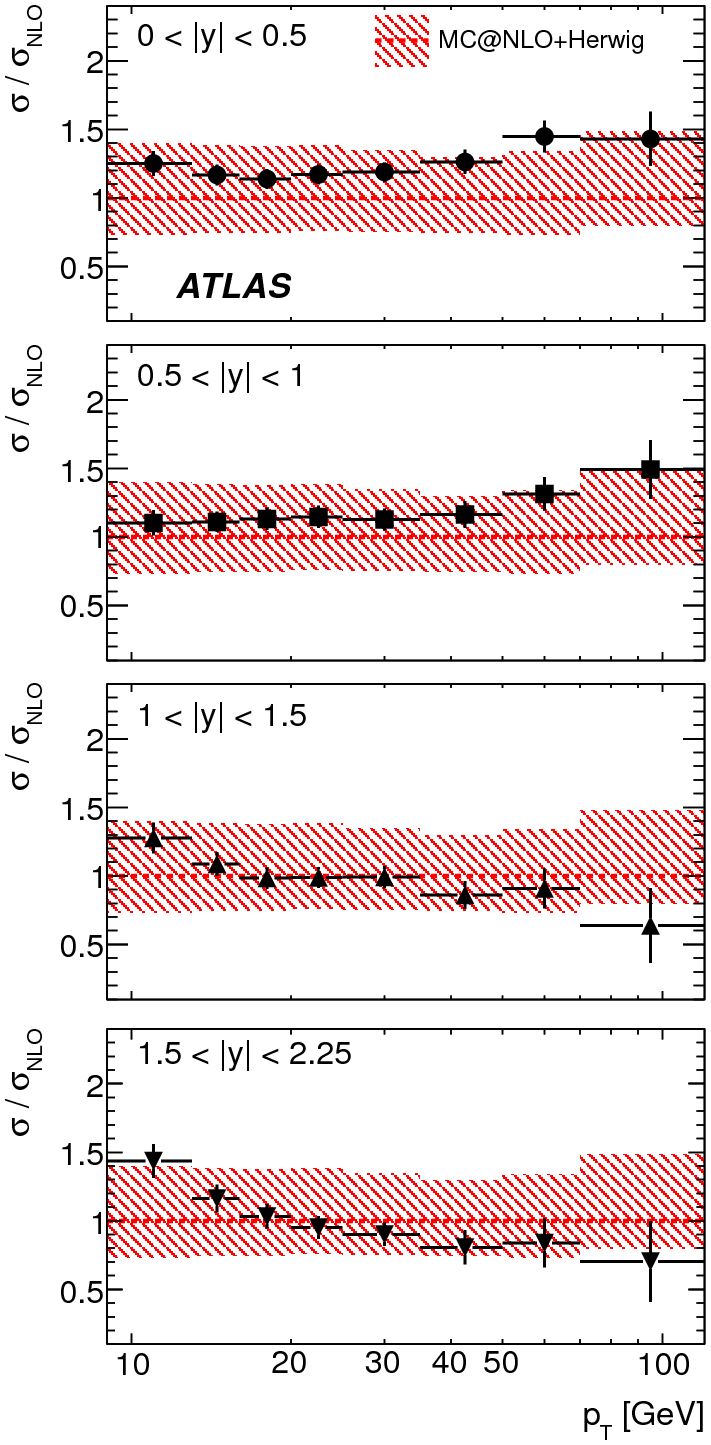}}
  \caption{ (a) The doubly-differential cross-section for $B^+$ production as a function of $p_T$ and $y$, averaged over each ($p_T$, $y$) interval and quoted at its center. The data points are compared to NLO predictions from POWHEG and MC@NLO. The shaded areas around the theoretical predictions reflect the uncertainty from renormalization and factorization scales and the b-quark mass. The ratio of the measured cross-section to the theoretical predictions ($\sigma/\sigma_{NLO}$) of POWHEG (b) and MC@NLO (c) in eight $p_T$ intervals in four rapidity ranges is shown. The points with error bars correspond to data with their associated uncertainties, which is the combination of the statistical and systematic uncertainty. The shaded areas around the theoretical predictions reflect the uncertainty from renormalization and factorization scales and the b-quark mass.}
  \label{fig:bpluscs}
\end{figure}
\begin{figure}[!htb]
  \centering
  \includegraphics[height=2in]{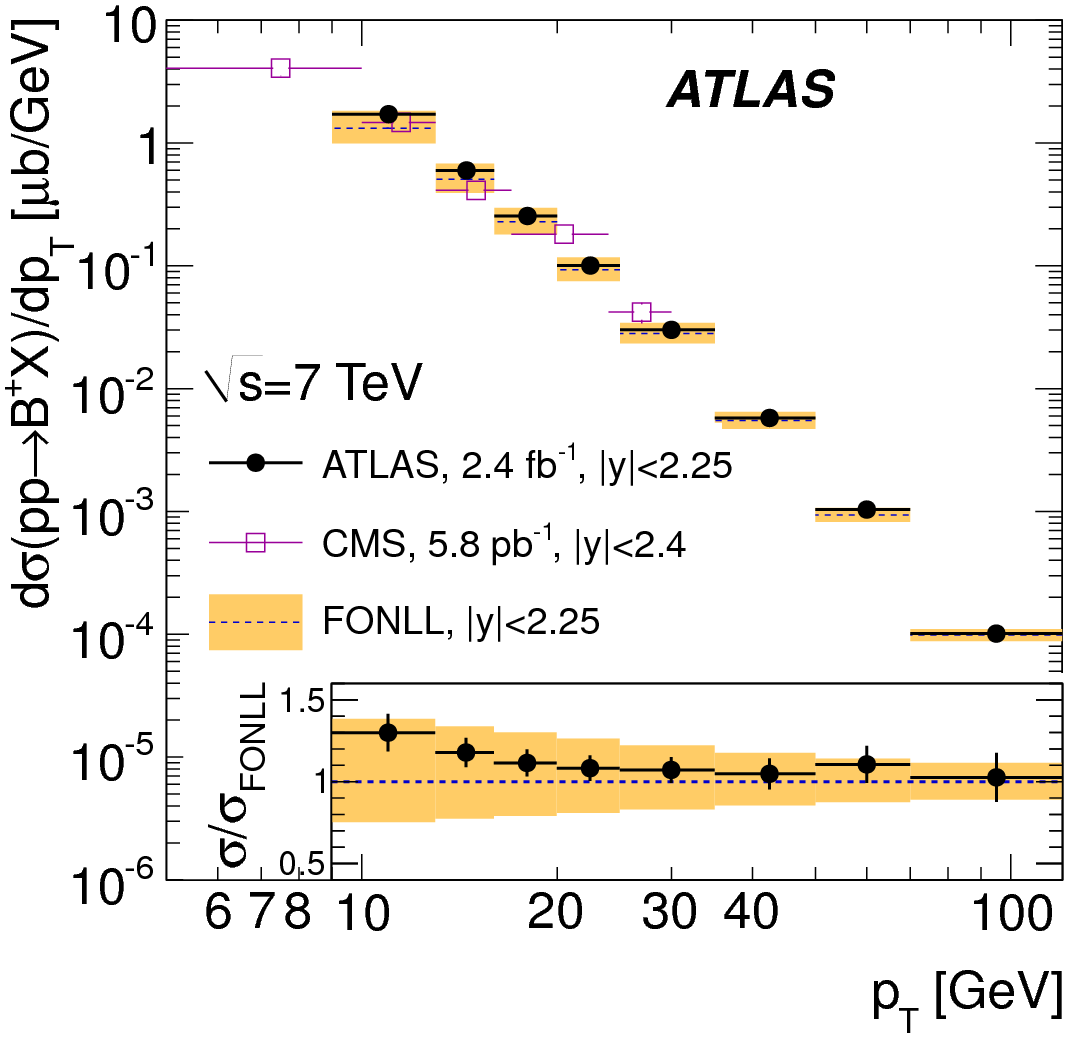}
  \caption{ The differential cross-section for $B^+$ production versus $p_T$, integrated over rapidity. The solid points with error bars correspond to the differential cross-section measurement by ATLAS with total uncertainty (statistical and systematic) in the rapidity range $|y|<$ 2.25, averaged over each $p_T$ interval and quoted at its center. For comparison, data points from CMS are also shown, for a measurement covering $p_T<$ 30 GeV and $|y|<$ 2.4 \cite{bplusCMS}. Predictions of the FONLL calculation for b-quark production are also compared with the data, assuming a hadronization fraction $f_{b\rightarrow B^+}$ of $(40.1\pm 0.8)\%$ \cite{bplusf} to fix the overall scale. Also shown is the ratio of the measured cross-section to the predictions by the FONLL calculation ($\sigma/\sigma_{FONLL}$). The upper and lower uncertainty limits on the prediction were obtained by considering scale and b-quark mass variations.}
  \label{fig:bpluscsFONLL}
\end{figure}

ATLAS has measured the $B^+$ production cross-section using the decay channel $B^+\rightarrow J/\psi K^+$ in pp collisions at $\sqrt{s}$ = 7 TeV, as a function of $B^+$ transverse momentum and rapidity \cite{bplus}. The measurement uses data of integrated luminosity of 2.4 $fb^{-1}$ recorded in early 2011 data $|y| < $ 2.25 and $p_T$ up to 100 GeV. Events were selected using a di-muon trigger where both muons are required to have $p_T > $ 4 GeV and pass a loose selection requirement compatible with $J/\psi$  meson decay into a muon pair. The $B^+$ candidates are reconstructed from a $J/\psi$ candidate combined with a hadron track with $p_T>$ 1 GeV and required to have $p_{T} >$ 9 GeV, $|y| <$ 2.3, and $\chi^2/N_{d.o.f}<$ 6. The muon tracks and hadron tracks are required to have sufficient numbers of hits in the Pixel, Semiconductor Tracker (SCT) and Transition Radiation Tracker (TRT) detectors to ensure accurate ID measurements.\\

The differential cross-section for $B^+$ production is given by\\
$$\frac{\mathrm{d}^2\sigma(pp\rightarrow B^+X)}{\mathrm{d}p_T\mathrm{d}y}\cdot{\cal{B}} = \frac{N^{B^{+}}}{{\L}\cdot\Delta p_{T} \cdot \Delta y},$$\\
where $N^{B^{+}}$ is derived from the average yield of the two reconstructed charged states $B^+$ and $B^-$ in each ($p_T$, $y$) interval after correcting for detector effects and acceptance. The total number of $B^\pm$ events is extracted using a binned maximum likelihood fit in each ($p_T$, $y$) bin (\eg~Figure~\ref{fig:bplusfit}). The differential cross-section is calculated in 8 $p_T$ bins and 4 $y$ bins in the full kinematic range 9 GeV $<p_T<$ 120 GeV and $|y|<$ 2.25 as shown in Figure~\ref{fig:bpluscs}(a). The data points have been compared to next-to-leading order (NLO) \cite{mb,NLO} predictions from POWHEG \cite{NLOP,NLOP2} and MC@NLO \cite{NLOM,NLOM2}. As shown in Figure~\ref{fig:bpluscs}(b)(c), the POWHEG prediction shows good agreement with data in both cross section and shape while MC@NLO predicts a lower cross section and softer $p_T$ spectrum at low $y$ and a harder $p_T$ spectrum in high $y$. The results have been compared to fixed order plus next-to-leading logarithms (FONLL) \cite{FONLL,FONLL2} as well. As shown in Figure~\ref{fig:bpluscsFONLL}, the FONLL prediction shows good agreement especially with $p_T<$ 30 GeV, and is still compatible in the high $p_T$ range even up to 100 GeV.

\section{$B_c$ meson observation}
The $B_{c}^\pm$ meson is a bound state of the two heave quarks able to form a stable state. Weak decays of the $B_{c}^\pm$ meson provide a unique probe of heavy quark dynamics that is inaccessible to $b\bar{b}$ or $c\bar{c}$ bound states. \\
\begin{figure}[!htb]
  \centering
  \includegraphics[height=2in]{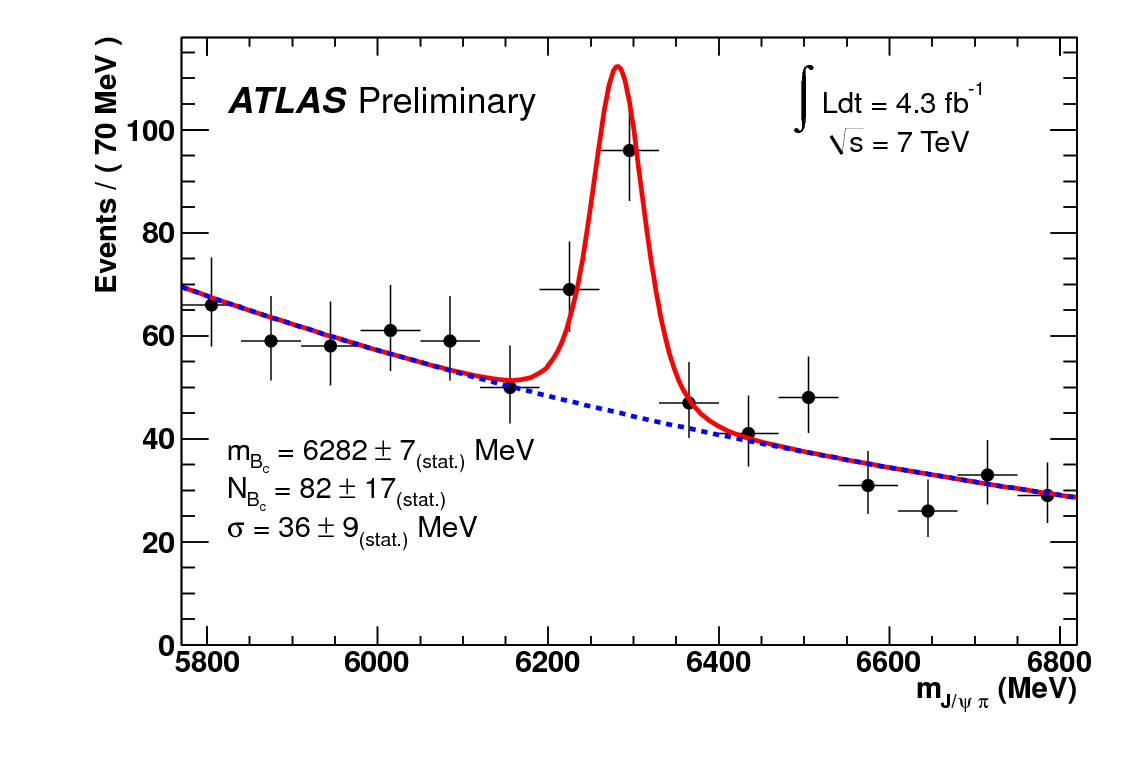}
  \caption{The invariant mass distribution of reconstructed $B_c^{\pm} \to J/\psi \pi^{\pm}$ candidates. The points with error bars are the data. The solid line is the projection of the results of the unbinned maximum likelihood fit to all candidates in the mass range 5770-6820 MeV. The dashed line is the projection for the background component on the same fit.}
  \label{fig:bc}
\end{figure}

The $B_{c}^\pm$ meson is reconstructed using ATLAS pp collision data collected in the year 2011 at $\sqrt{s}$ = 7 TeV using the decay mode $B_c\rightarrow J\psi\pi$ \cite{bc}. Single and di-muon triggers with various transverse momentum have been used. Each muon candidate must have a track reconstructed in the MS combined with a track reconstructed in the ID. The di-muon selection requires a pair of oppositely charged muons with $p_T>$ 4 GeV for both in the first half year and $p_T>$ (6, 4) GeV in the second half year. Additionally they are required to pass a loose selection requirement compatible with $J/\psi$  meson decay into a muon pair. The $B_c$ candidate is reconstructed from the $J/\psi$ candidate combined with a hadron track with $p_T>$ 4 GeV. Unlike $B^+$, the short $B_c$ meson lifetime means that lifetime cuts are not efficient in separating the $B_c$ signal from direct $J/\psi$ combinations. The transverse impact parameter significance has been proven to be more efficient, and it is required to be larger than 5. The reconstructed $B_c$ candidates are required to have $p_{T} >$ 15 GeV and $\chi^2/N_{d.o.f}<$ 2.\\

With the full 2011 integrated luminosity of 4.3 $fb^{-1}$ of data, $82\pm17$ $B_c$ ground state mesons have been extracted using an unbinned maximum likelihood fit (Figure~\ref{fig:bc}). The $B_c$ mass returned by the fit is $6282\pm7 $ MeV, which is consistent with the PDG mass\cite{bcPDG} of $6277\pm7$ MeV.
\section{Measurement of the $\Lambda_b$ lifetime and mass}
$\Lambda_{b}^0$ is the lightest b-baryon. Although it has been measured by many experiments \cite{lambdab2, lambdab3,lambdab4}, its lifetime still has large experimental uncertainty (2-3\%), and the discrepancy between the CDF and D0 measurements is high (1.8$\sigma$).\\
\begin{figure}[!htb]
  \centering
        \subfigure[]{\label{a}\includegraphics[height=2in]{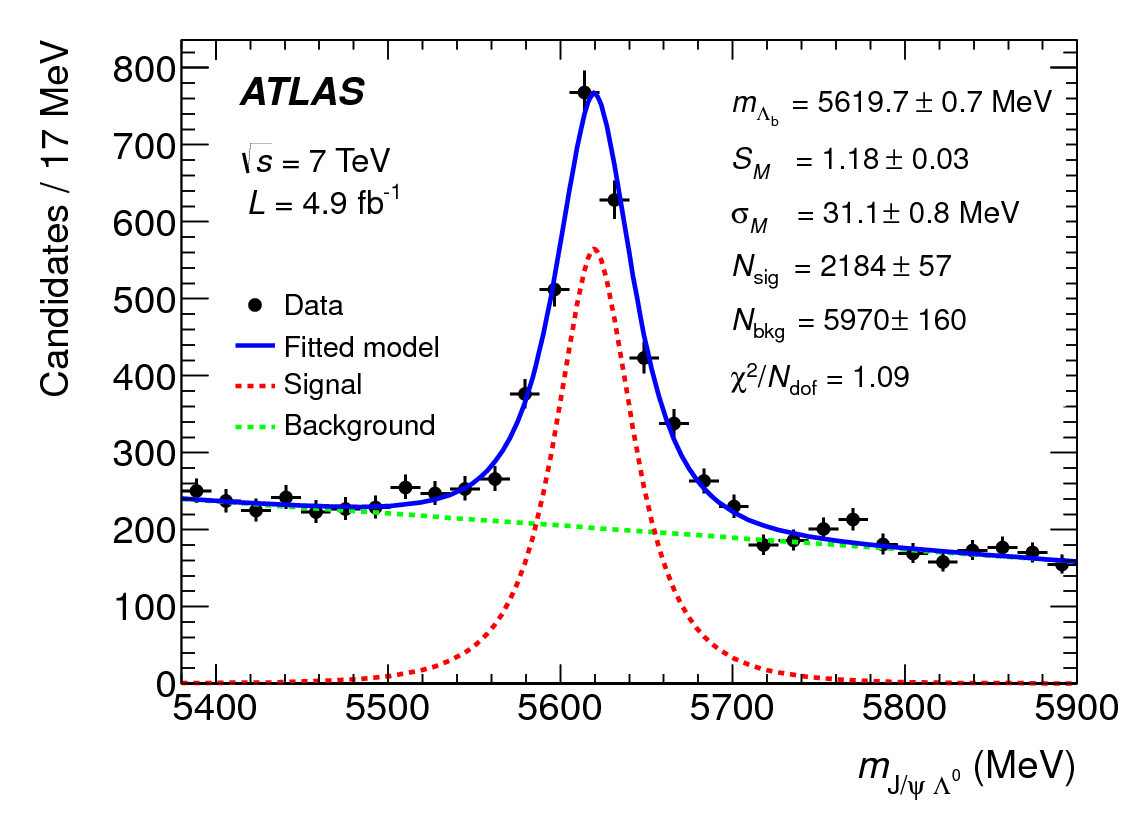}}
    	\subfigure[]{\label{b}\includegraphics[height=2in]{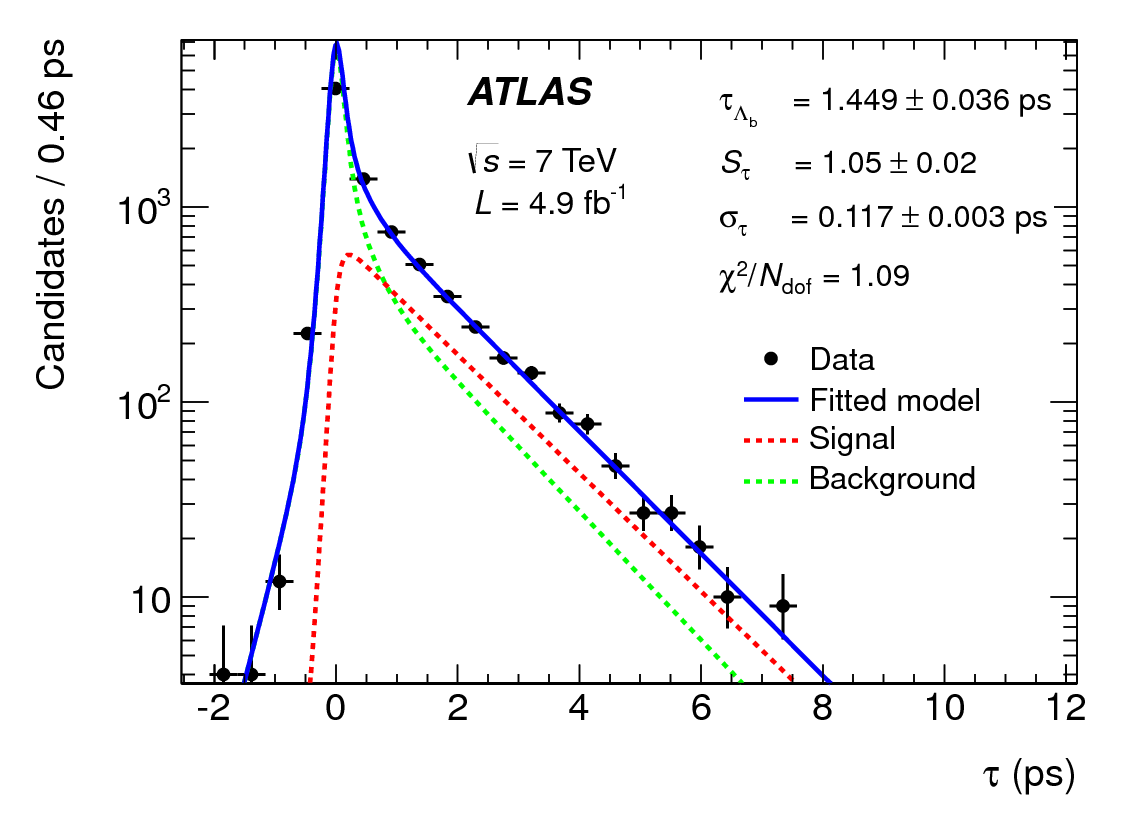}}
  \caption{ Projection of the fitted probability density function onto the mass (a) and the proper decay time (b) axis for $\Lambda_{b}$ candidates. The displayed errors are statistical only. The $\chi^2/N_{d.o.f}$ value is calculated from the dataset binned in mass and decay time with the number of degrees of freedom $N_{d.o.f}$ = 61.}
  \label{fig:lambdab}
\end{figure}

This analysis is based on 4.9 $fb^{-1}$ data collected by ATLAS in 2011 using single muon and di-muon $J/\psi$ triggers and reconstructed through the decay channel of $\Lambda_b^0\rightarrow J/\psi(\mu^+\mu^-)\Lambda^0,~ \Lambda^0\rightarrow p^+\pi^-$ \cite{lambdab}. Muon tracks are reconstructed in the ID and identified in the MS, with $p_T$ above 400 MeV and pseudorapidity $|\eta| < $ 2.5 required. The reconstructed $J/\psi$ candidates are selected within a mass window of 2.8 GeV $<m(\mu^+\mu^-)<$ 3.4 GeV with a vertex fit constrained to the $J/\psi$ mass of 3069.92 MeV and muon $p_T>$ 4 GeV. Reconstructed $\Lambda^0$ candidates are selected within a mass window of 1.08 GeV $<m(p\pi)<$ 1.15 GeV with a vertex fit constrained to the $\Lambda^0$ mass of 1115.68 MeV; additionally the $\Lambda^0$ vertex is required to point to the $J/\psi$ vertex. Finally the $\Lambda_{b}^0$ candidates are reconstructed using a cascade vertex fit applied to the four tracks $\mu^+,~\mu^-,~p,~\pi$ simultaneously with all the $J/\psi$ and $\Lambda^0$ constraints. For comparison, $B_d$ candidates with similar topology are reconstructed as well. The reconstructed $\Lambda_{b}^0$ candidates are required to have $p_T>$ 3.5 GeV, transverse decay length larger than 10 mm, global $\chi^2/N_{d.o.f}<$ 3, and the difference between the cumulative $\chi^2$ probabilities of the two fits larger than 0.05.\\

The mass and the proper decay time, defined as:
$$\tau=\frac{L_{xy}m^{PDG}}{p_T}$$
 of the $\Lambda_{b}^0$ are extracted using a simultaneous unbinned maximum likelihood fit with per event error (Figure~\ref{fig:lambdab}). Including systematic uncertainties on event selection and reconstruction bias,  the fit model, $B_d^0$ contamination, and the $p_T$ scale, the mass and proper decay time measured are:
 $$m_{\Lambda_b^0} = 5619.7 \pm 0.7(stat) \pm 1.1(syst)~\rm{MeV}$$
 $$\tau_{\Lambda_b^0} = 1.449 \pm0.036(stat)\pm0.017(syst)~\rm{ps}.$$
 They are consistent with the PDG value \cite{bplusf} and the LHCb result \cite{lambdabLHCb}.\\

 The ratio of the $\Lambda_{b}^0$ and $B_{d}^0$ lifetimes has been measured:
$$R = \tau_{\Lambda_b^0}/\tau_{B_d^0} = 0.960 \pm 0.025(stat) \pm 0.016(syst).$$
This ratio is intermediate to the recent determination by D0, $R^{D0} = 0.864 \pm 0.052(stat)\\ \pm 0.033(syst)$ \cite{lambdabD0}, and the measurement by CDF, $R^{CDF} = 1.020 \pm 0.030(stat) \pm
0.008(syst)$ \cite{lambdabCDF}. It agrees with heavy quark expansion calculations which predict the value of the ratio to be between 0.88 and 0.97 \cite{hqe} and is compatible with the next-to-leading order QCD predictions with central values ranging between 0.86 and 0.88 (uncertainty of $\pm~0.05$) \cite{NLOQCD}.
\section{Observation of a new $\chi_{b}$ state in radiative transitions to $\Upsilon(1S)$ and $\Upsilon(2S)$}

\begin{figure}[!htb]
  \centering
  \includegraphics[height=3in]{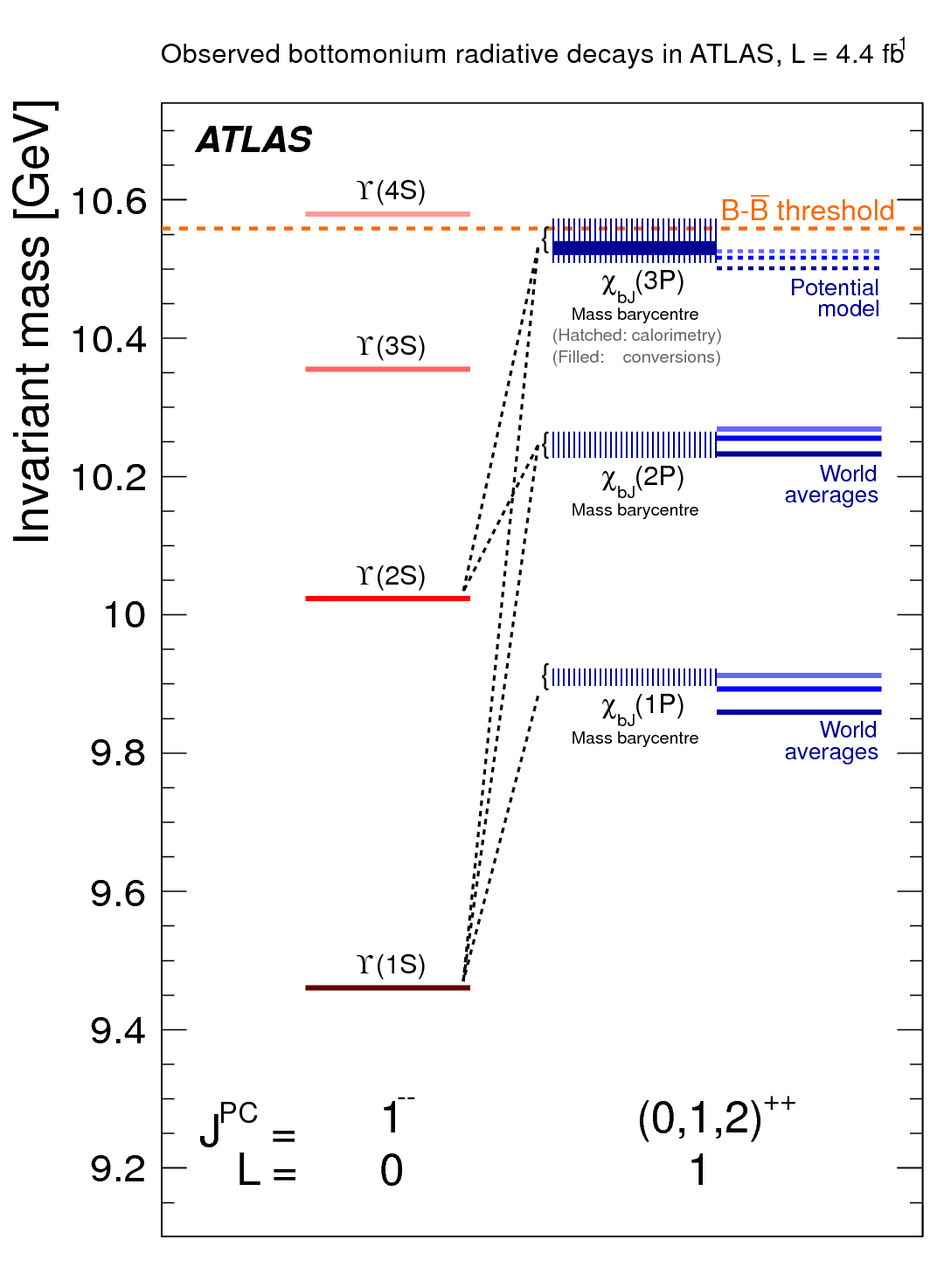}
  \caption{ Masses and radiative transitions of the $\chi_{b}$ states as observed in ATLAS. The first column shows the world-average masses of the Υ states. The second column displays the ATLAS mass barycenter determination for each of the $\chi_{b}(1P,2P,3P)$ triplets, with the bands of vertical stripes indicating the measurements from the analysis using unconverted photons and the solid band representing the measurement from the converted photon analysis ($\chi_{b}(3P)$ only). Dashed lines between the states in the second and first columns indicate the radiative transitions observed by ATLAS. In the third column, the world averages (or Potential Model predictions in the case of the 3P) are shown for the three multiplets. Progressively paler shades of blue indicate the J = 0, 1, 2 states.}
  \label{fig:chibpredict}
\end{figure}
\begin{figure}[!htb]
  \centering
  \includegraphics[height=2in]{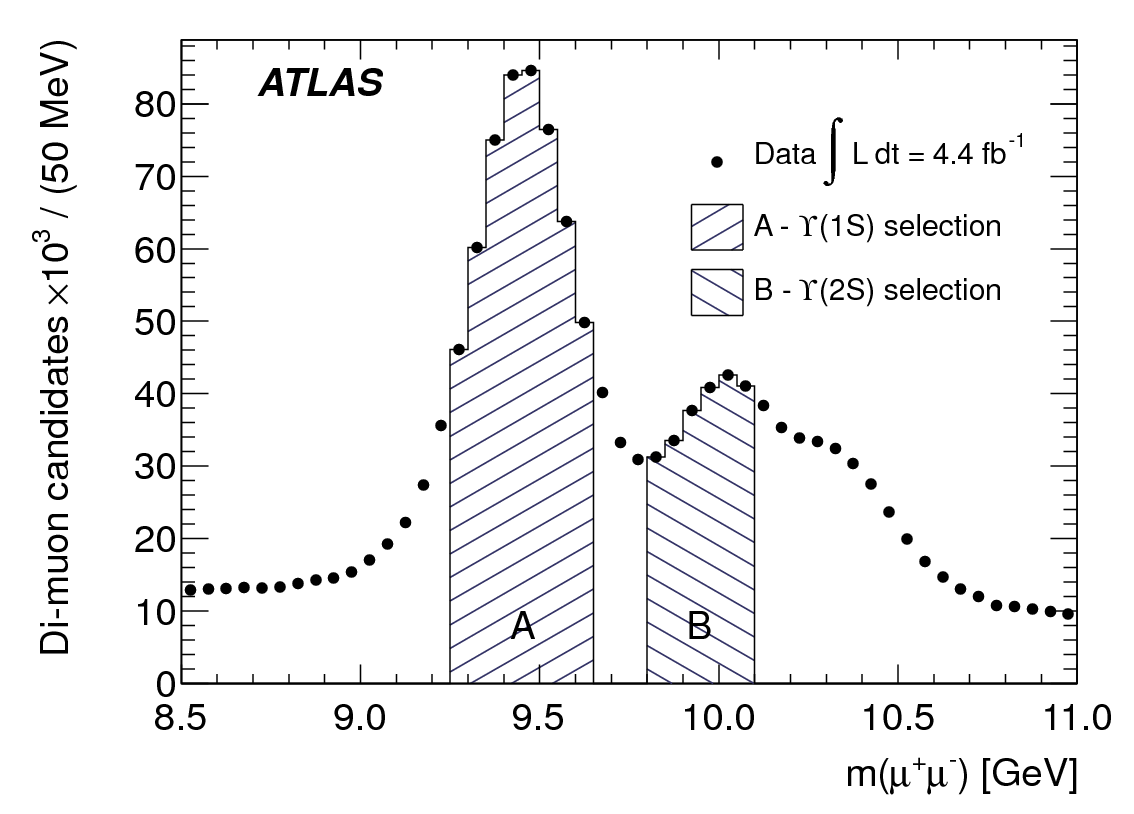}
  \caption{ The invariant mass of selected di-muon candidates. The shaded regions A and B show the selections for $\Upsilon(1S)$ and $\Upsilon(2S)$ candidates respectively.}
  \label{fig:chibmass}
\end{figure}

The $\chi_{b}(1P)$ and $\chi_{b}(2P)$ states have been observed in previous experiments, but the $\chi_{b}(3P)$ state has not. The $\chi_{b}(3P)$ state is the highest P state predicted below the $B-\bar{B}$ threshold as shown in Figure~\ref{fig:chibpredict} with mass about 10.52 GeV and hyperfine splitting about 10-20 MeV. The $\chi_{b}(nP)$ states are sought using a data sample corresponding to an integrated luminosity of 4.9 $fb^{-1}$ of ATLAS 2011 data \cite{chib}, through decay modes of $\chi_{b}(nP)\rightarrow\Upsilon(1S)\gamma$ and $\chi_{b}(nP)\rightarrow\Upsilon(2S)\gamma$. \\
\begin{figure}[htb]
  \centering
    	\subfigure[]{\label{a}\includegraphics[height=2in]{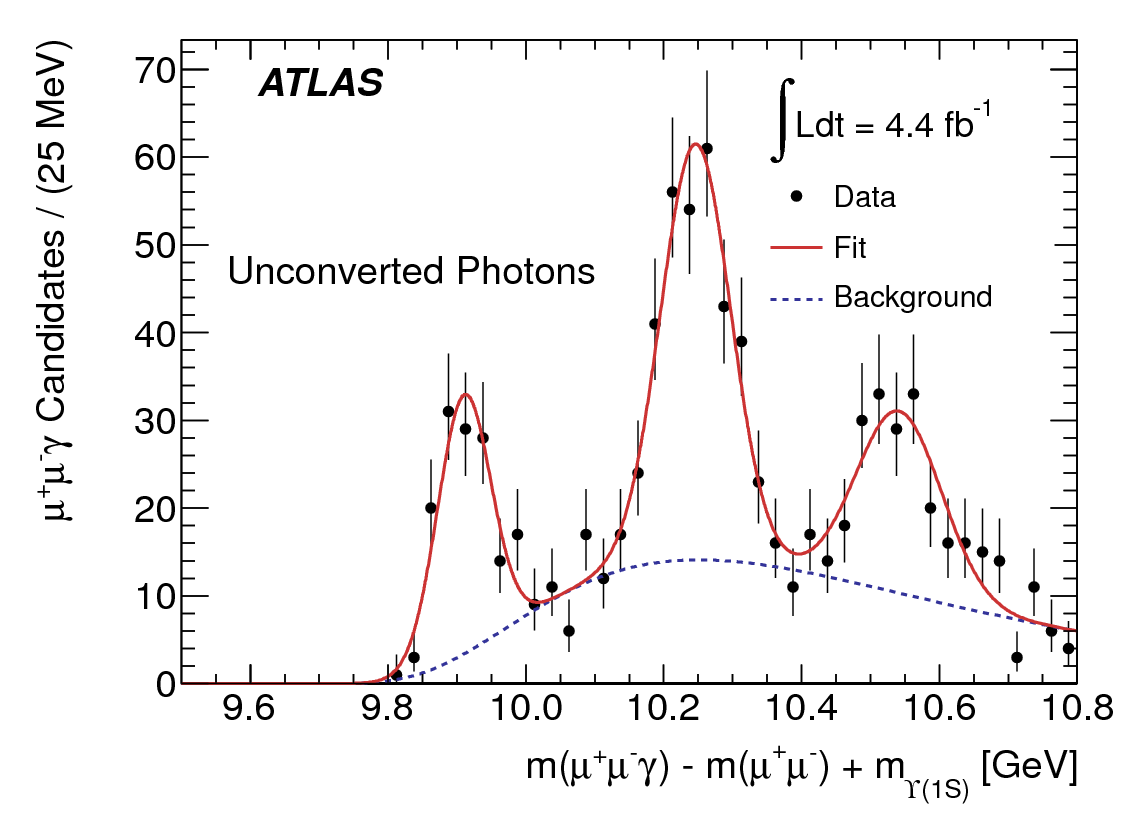}}
    	\subfigure[]{\label{b}\includegraphics[height=2in]{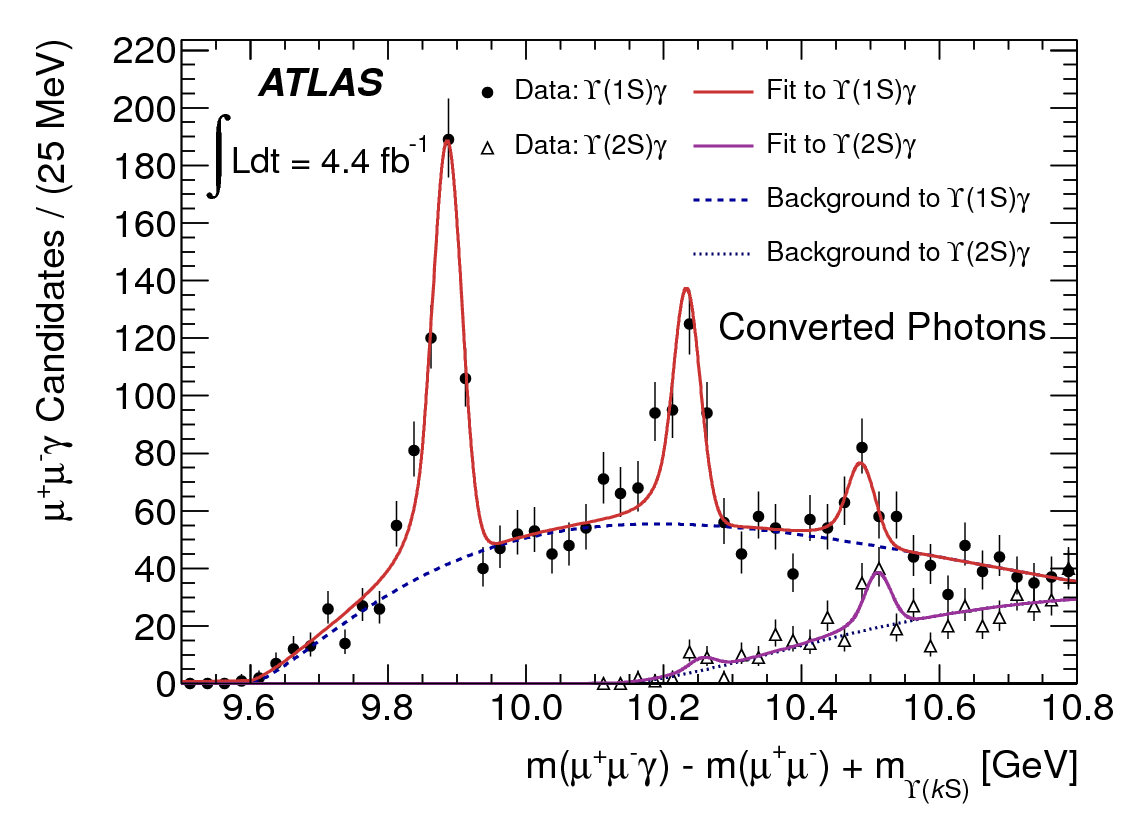}}
  \caption{ (a) The mass distribution of $\chi_{b}\rightarrow\Upsilon(1S)\gamma$ candidates for unconverted photons reconstructed from energy deposits in the electromagnetic calorimeter ($\chi^2_{fit}/N_{d.o.f.}$ = 0.85). (b) The mass distributions of $\chi_{b}\rightarrow\Upsilon(kS)\gamma$ (k = 1, 2) candidates formed using photons which have converted and been reconstructed in the ID ($\chi^2_{fit}/N_{d.o.f.}$ = 1.3). Data are shown before the correction for the energy loss from the photon conversion electrons due to bremsstrahlung and other processes. The data for decays of $\chi_{b}\rightarrow\Upsilon(1S)\gamma$ and $\chi_{b}\rightarrow\Upsilon(2S)\gamma$ are plotted using circles and triangles respectively. Solid lines represent the total fit result for each mass window. The dashed lines represent the background components only.}
  \label{fig:upsilon}
\end{figure}

In this measurement, a set of muon triggers designed to select events containing muon pairs or single high transverse momentum muons was used to collect the data sample. Each muon candidate must have a track reconstructed in the MS combined with a track reconstructed in the ID with $p_{T} >$ 4 GeV and pseudorapidity $|\eta| <$ 2.3. The di-muon selection requires a pair of oppositely charged muons which are fitted to a common vertex. The di-muon candidate is also required to have $p_{T} >$ 12 GeV and $|\eta| <$ 2.0. $\Upsilon(1S)\rightarrow\mu\mu$ candidates with masses in the range $9.25 < m_{\mu\mu} < 9.65$ GeV and $\Upsilon(2S)\rightarrow\mu\mu$ candidates with masses in the range $9.80 < m_{\mu\mu} < 10.10$ GeV are selected (Figure~\ref{fig:chibmass}). This asymmetric mass window for $\Upsilon(2S)$ candidates is chosen in order to reduce contamination from the $\Upsilon(3S)$ peak and continuum background contributions. A photon is combined with each $\Upsilon$ candidate. Converted photons reconstructed by ID tracks from $e^+e^-$ pairs with a conversion vertex and unconverted photons reconstructed by electromagnetic calorimeter energy deposit are used. The converted photon candidates are required to be within $|\eta| <$ 2.30 while the unconverted photon candidates are required to be within $|\eta| <$ 2.37. Unconverted photons must also be outside the transition region between the barrel and endcap calorimeters, 1.37 $< |\eta| <$ 1.52. Requirements of $p_{T}(\mu^+\mu^-) >$ 20 GeV and $p_{T}(\mu^+\mu^-) >$ 12 GeV are applied to $\Upsilon$ candidates with unconverted and converted photon candidates respectively. These thresholds are chosen in order to optimize signal significance in the $\chi_{b}(1P,2P)$ peaks.\\

As shown in the mass difference $m(\mu^+\mu^-\gamma)-m(\mu^+\mu^-)+m_{PDG}(\Upsilon)$ distributions (Figure~\ref{fig:upsilon}), in addition to the mass peaks corresponding to the decay modes of $\chi_{b}(1P,2P)\rightarrow\Upsilon(1S,2S)$, a new structure centered at mass $10.530 \pm 0.005 (stat.) \pm 0.009 (syst.)$ GeV is observed with a significance of more than 6$\sigma$ in both of those two independent samples. This is interpreted as the $\chi_{b}(3P)$ state.\\


\end{document}